\documentclass[letterpaper,11pt]{article}
\usepackage[T1]{fontenc}
\usepackage{graphicx}
\usepackage{url}
\usepackage{rotating}
\usepackage{latexsym}

\begin{document}

\title{Mapping Quantum Circuits to Ions in Storage Ring Quantum Computer Architectures}
\author{Kevin Brown\thanks{Brookhaven National Laboratories, Upton NY 11973, 
	United States}
\and 
Thomas Robertazzi\thanks{ECE Dept., (affiliate) Applied Mathematics and Statistics Dept., Stony Brook University, NY 11794,
Stony Brook, New York, United States.}
\footnote{%
Corresponding author.} 
}
\date{\today}

\maketitle 

\begin{abstract}

The mapping of quantum circuits to qubits represented by ion states in a storage ring quantum 
computer is examined.  Serial, parallel and hybrid architectures for mapping such qubits 
onto a storage ring quantum computer are presented.  The timing of laser actions and timing equations are discussed.  A \lq\lq predecessor problem" 
that arises in such architectures is identified and a solution is proposed.  Representative numerical sizing and timing calculations are presented.  The entire 
methodology is very general and extends to several implementation options.  Some open problems are identified.  

\end{abstract}


\section{Introduction}

This paper presents architectural means of implementing quantum circuits on a storage ring quantum computer.  A storage ring quantum computer \cite{Review, Patent-1} is a particle accelerator (generally circular in form, although with straight sections it can take on other shapes, including triangular, square, hexagonal, etc.) that uses circulating ions to store quantum qubit information.  The proposed size of such a storage ring is on the order of  magnitude of one meter in circumference.  Quantum computing operations can be performed on the circulating ions$/$qubits.  Vacuum chamber windows on the storage ring allow lasers to interact with the ions to perform quantum computing operations.  
A qubit, in a storage ring quantum computer, is either an internal quantum eigenstate of the individual ions or is an external quantum eigenstate of a chain of ions forming an ion Coulomb crystal. We assume that a system is in place (i.e., lasers to excite quantum states) to write and measure these qubits and using this technology we consider how to implement quantum gate circuits in the storage ring quantum computer.

Non-circular, linear, ion traps have long been implemented to use ions to store quantum information and perform quantum computing operations with the ions.  The implementation of a storage ring quantum computer is substantially different from a linear ion trap in that the ions rotate away from a point on the storage ring with time while the ions in a linear ion trap are stationary.  This brings new challenges and open problems in engineering a working SRQC.  Storage rings have been constructed in the past but not for computational purposes to date.  However there is much potential in this concept.  A SRQC can accommodate thousands of ions which are long lasting (on the order of minutes).  This is significantly larger than the number of qubits that can be currently implemented with other technologies.  There is also much that can be done in terms of the parallelism of the ions, windows and lasers.    

In section 2 storage ring quantum computer technology is reviewed.  Quantum circuit basics are discussed in section 3.  A method for implementing large depth quantum circuits on a SRQC is introduced in section 4.  The serial,  parallel and hybrid modes of SRQC operation are described in section 5.  Timing equations for the serial, parallel and hybrid modes of operation are presented in section 6.  Mismatched inputs and outputs are discussed in section 7.  The predecessor problem and  a solution appear in section 8. Some potential extensions of the technology are reviewed in section 9.  Representative numerical timing and sizing calculations are presented in section 10.   The conclusion appears in section 11.  

\section{Storage Ring Quantum Computers}

A storage ring quantum computer (SRQC) is built on the concept of a well established particle accelerator device called a circular 
radio frequency quadrapole (CRFQ).  A CRFQ may be viewed as an unbounded Paul trap.  A CRFQ need not be large.  One preliminary design is a device 
one meter in circumference that can hold thousands of ions.    

For quantum mechanical phenomena to be seen in a classical particle beam, the beam must be cooled to very low temperatures \cite{Schatz}.  There are two fundamental states of matter that can be created by bringing a beam's temperature to a very low value:

\vspace{0.15 in}

$\bullet$ Classical crystalline beam \cite{Schatz}: A cluster of circulating, charged particles in its classical lower energy state influenced by 
circumferentially varying guiding and focusing electro-magnetic forces and Coulomb interacting forces \cite{Wei}.  

\vspace{0.15 in}

$\bullet$ Ultracold crystalline beam: The second state of matter is an ultracold crystalline beam which will be referred to as an ion Coulomb 
crystal.  This state of matter is cooled to well below the Doppler cooling limit to the resolved sideband limit 
but not as low as the Lamb-Dicke limit \cite{Steane, Wineland}.  This is a \lq\lq Goldilocks" arrangement, where couplings between internal and external quantum states are not excessively suppressed and is far above the temperature for a Bose-Einstein condensate \cite{Steane}. 

\vspace{0.15 in}

A number of quantum properties have the potential to be taken advantage of in a classical crystalline beam using standard measurement methods.  These include the crystalline orbit 
modes, the spin states of particles and emittance quantization.  However these properties are not controllable because of the high temperatures of the beams.  They are thus not of practical use for computation.  An ultracold crystalline beam operates at low enough temperatures that quantum mechanical properties can be accessed for use in computational processes.

It should be noted that the existing quantum computer architecture closest to the SRQC concept is that of ion traps.  An ion trap confines ions using a transverse radio-frequency electric field and electrostatic end potentials.   Ion traps have ions either in stationary positions 
\cite{Linke, Murali}, or the ions can be shuttled 
about \cite{Kiel, Saki} or they can form a linear arrangement that can be moved back and forth under a reading/writing device \cite{Wu}.

Ion trap research has demonstrated that quantum states in trapped ions can be stable for very long times, including the scale of 
minutes \cite{Wineland-2}. This is promising for ions in a storage ring.  But note that the significant issues for ion traps are a concern for storage rings.  These include spontaneous transitions in the vibrational motion, thermal radiation and instabilities in systems such as rf voltage, laser power and mechanical vibrations.

In a  storage ring, groups of ions can be isolated from each other by means of longitudal rf potentials, or by separation thru the modulation of the velocity in the cooling system.  This can create isolated groups of ions holding specific qubit information that can be operated on independently.  The 
concept of a multiplex environment in linear ion traps is discussed in \cite{Kiel, Tabakov}.  Under multiplexing a storage ring could hold thousands of smaller individual crystals.  Such numerous small chains could be used for purposes such as quantum circuit implementation, systematic analysis or quantum memory. 

See \cite{Review} for an expanded treatment of the topic of this section. 

\section{Quantum Circuits}

Quantum circuits are interconnections of quantum gates that can accomplish some desired overall function \cite{Bernhardt, Deutsch, DiVincenzo,  Nielsen, Roloff}.  There are well known quantum circuits for purposes such as teleportation, super dense coding and error correction.  Quantum gates include well known ones such as the H, CNOT, SWAP, I, Z, X and Y gates.

It is important to understand that at a fundamental level for a storage ring quantum computer, 
a quantum gate operating on a set of qubit inputs leaves the \lq\lq output qubits" in the same 
locations (or subset/super-set of locations) as the original input qubits.  Thus on a SRQC it is not that we are implementing  a traditional 
electric-like circuit with physically distinct gates that can be pointed to in a physical circuit, taking inputs at one location and producing outputs at different location(s). Rather 
the quantum circuit diagram indicates the pattern and timing of operations to be performed on ions with output qubits generally (but not always) in the 
same location(s) as input qubits.

Figure 1 illustrates a typical overall layout of a quantum circuit as a rectangle with inputs at the left and outputs at the right.  Internally the 
quantum circuity consists of an interconnection of quantum gates which one may view as being arranged in \lq\lq stages".  Using a convention appearing in 
the literature \cite{Kaye} the \lq\lq width" of the circuit is the number of inputs or (parallel) wires and the \lq\lq depth" of the circuit is the number of stages (see the figure).  Alternately consider the circuit as being divided into a number of discrete time slices where the application of a single gate involves a single time slice. Then the depth of the circuit is defined as the number of time slices from when inputs are presented to the circuit to 
when circuit outputs appear.  Since gates may operate in parallel, the number of time slices usually does not equal the number of gates in the circuit.

\begin{figure}[t]
	\centerline{\includegraphics[width=11cm]{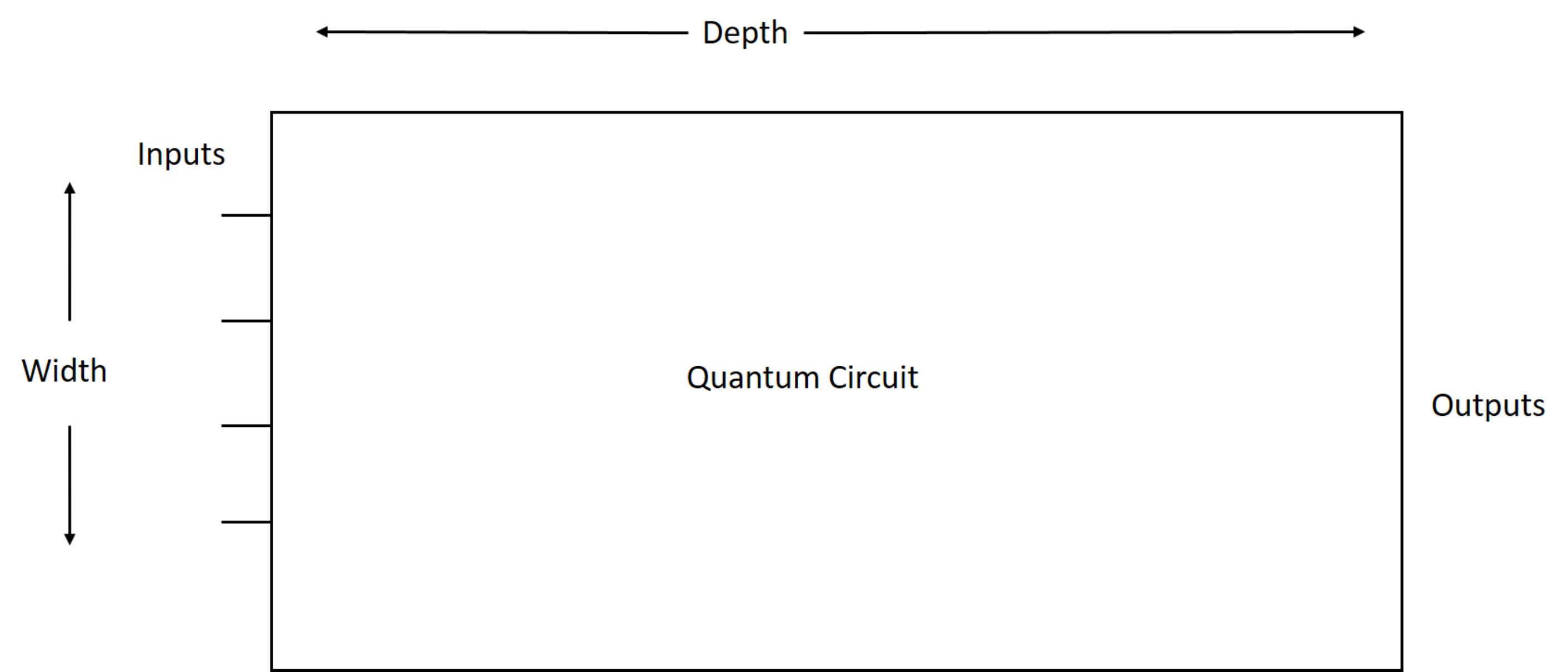}}
	\caption{Quantum Circuit Layout.}
	\label{fig-quantum-circuit}
\end{figure} 

Sometimes a \lq\lq gate" may be expressed as an equivalent interconnection of more basic gates \cite{Bae}.  Sometimes gates entangle multiple qubits.  Finally each 
logical qubit may be represented as a number of physical qubits (i.e ions and their states) \cite{Kiel}.  The following 
sections are written to cover all of these situations.       

In an ion trap system (and the SRQC) there is a three step process \cite{Murali}  in going from a quantum circuit with traditional quantum gates to a mapping of a circuit to the ions:

(1) Start with a tradtional quantum circuit that we wish to implement (with gates like CNOT, H etc..).

(2) Transform this circuit into an equivalent circuit using only (usually two) \lq\lq native" or \lq\lq hardware" gates.  These native gates are usually a one qubit gate and  a two qubit gate that can be implemented naturally in a linear ion trap or the SRQC.

(3) Map the operations corresponding to the native gate circuit to the ions.

One way to perform the laser operations corresponding to native gates is to use windows built into the storage ring.  Windows thru which lasers can access the ions which can be placed equally spaced around the ring or unequally spaced (some closer to one another than equi-spacing would allow) to minimize the travel time between windows for a faster multi-window calculation. One or more lasers may access ions thru a window.  Multiple 
lasers using a single window for access to ions/qubits are referred to as a \lq\lq laser bank" here.  Acoustic-optical modulators could be used to spatially steer each laser beam.  That is the acoustic-optical modulator can point a laser beam at an ion during part of all the ion's transit past the window.  The details of engineering multiple laser banks is an open problem.  This is one of several open problems pointed out in this paper.

\begin{figure}[t]
	\centerline{\includegraphics[width=14cm, height=10cm]{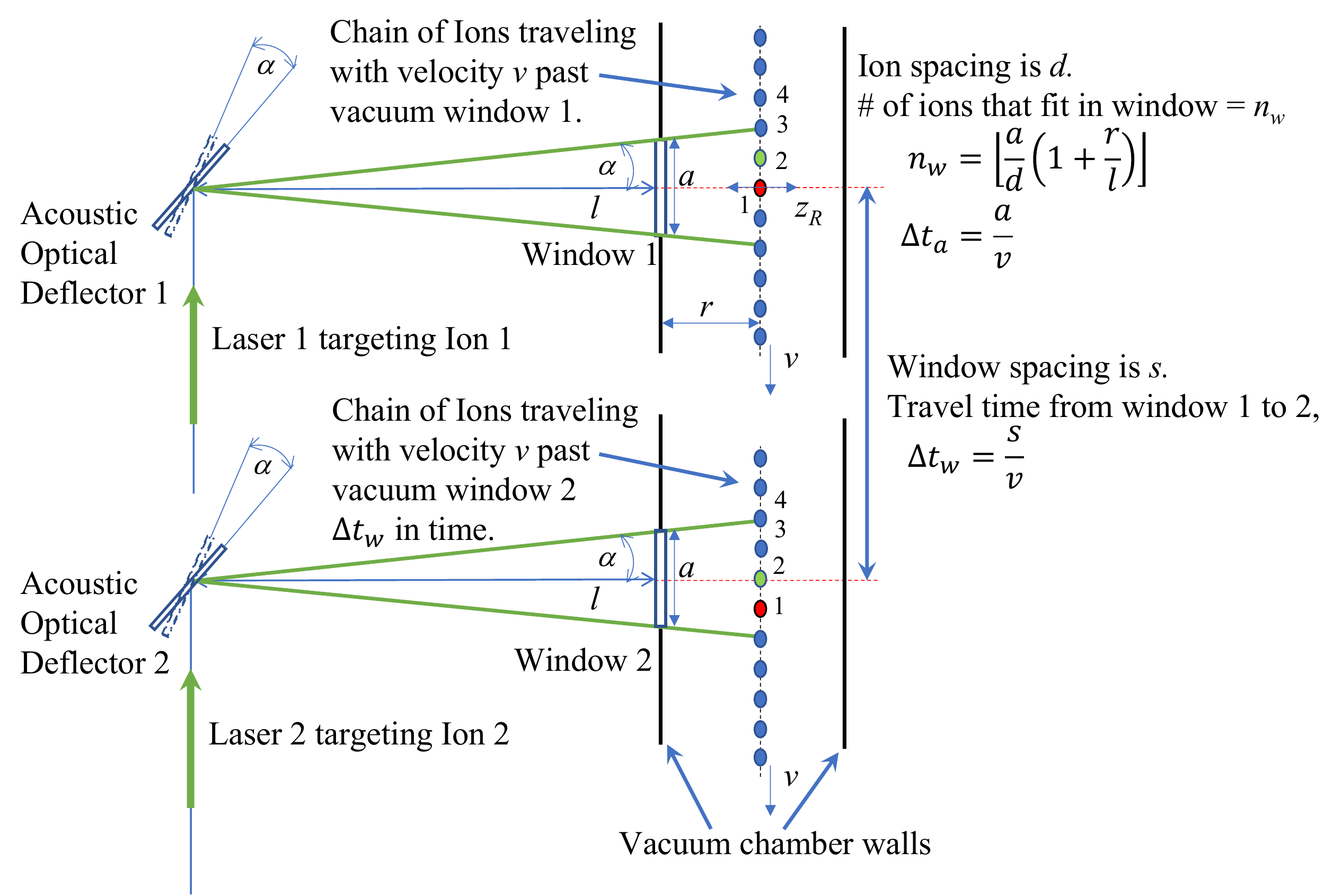}}
	\caption{Laser Programming Ions.}
	\label{fig-laser-timing}
\end{figure} 

An example is illustrated in Figure 2.  Ions 1 and 2 receive laser attention.  Some basic parameters and relationships 
are listed in the figure.  

\section{Implementing Large Depth Quantum Circuits}

We assume: 

\vspace{0.15 in}

$\bullet$ Assumption 1: For any gates, the output(s) appear at the end of a gate operation in the same or subset of the same qubit location(s) as the input(s).

\vspace{0.15 in}

Now let a certain number of consecutive qubits (width) come in view of a window (lasers) on the storage ring.  It will be accessible to the lasers for a certain amount of time until it rotates away.  During this time the lasers in the window can do gate operations on the qubits.  Multiple depth-wise stages 
of the circuit may be implemented during this time.  
But the depth of a circuit to be implemented for those qubits by the laser bank is limited by the time under the window$/$lasers.  
One needs to address the problem that occurs if the time the qubits are accessible to a (set of) laser(s) does limit the size of circuit depth/stages that can be implemented.    The method proposed here consists of the following:

For a large depth circuit one can do a number of operations/stages in the limited time window with the first set of lasers and then let the (stable) qubit ions rotate to a succeeding set of lasers (be they in the same window or a succeeding window) to continue operations.  This could be done multiple times, each time extending the depth of the circuit implemented.   This technique can be used to implement cascades of different quantum circuits which overall would constitute a very large depth quantum \lq\lq circuit".  Note that the width of the circuit (number of qubits for the circuit) here is limited by the number of lasers at one location and/or the size of the window.  

Implementing larger width circuits is discussed in section 5 where two modes of operation, serial and parallel, are introduced.  Timing equations which apply to both the serial and parallel modes of operation appear in section 6.

\section{Implementing Large Width Quantum Circuits}

Suppose a window has one or more banks of L lasers each and a laser bank can program L ion qubits, while they are in range of the laser bank, at a time.  Thus we could then handle a L qubit width or less circuit.  But could we handle a 2L qubit width or greater circuit?  We could think of having laser banks operate on different parts of an overall circuit (i.e. sub-circuits).

\vspace{0.10 in}

Some architectural possibilities are:

\vspace{0.1 in}

{\bf (A)Serial Mode of Operation:} Process the first L qubits for the first or more than one stage(s) (depth wise) of circuit.  Then let these rotate away to the next laser bank/window, bringing into view the next L ion/qubits to be processed by the first laser bank/window.  The original L qubits can be further processed by the next (2nd) laser bank/window.  

Figure 3 illustrates this.  Ions move from left to right in the figure.  Three laser banks are shown though a window may have anywhere from one laser to multiple laser banks depending on its size and configuration.  Ions are divided into bunches of L ions each.  Ions are uniformly spaced 
around the storage ring.  As an example, in the figure the width 
W is shown to be 4L.  That is W=4L.

With some thought one can see that serial mode is most suited for a circuit with triangular topology or initial triangular topology moving in through the circuit in a depth wise sense.

\begin{figure}[t]
	\centerline{\includegraphics[width=11cm]{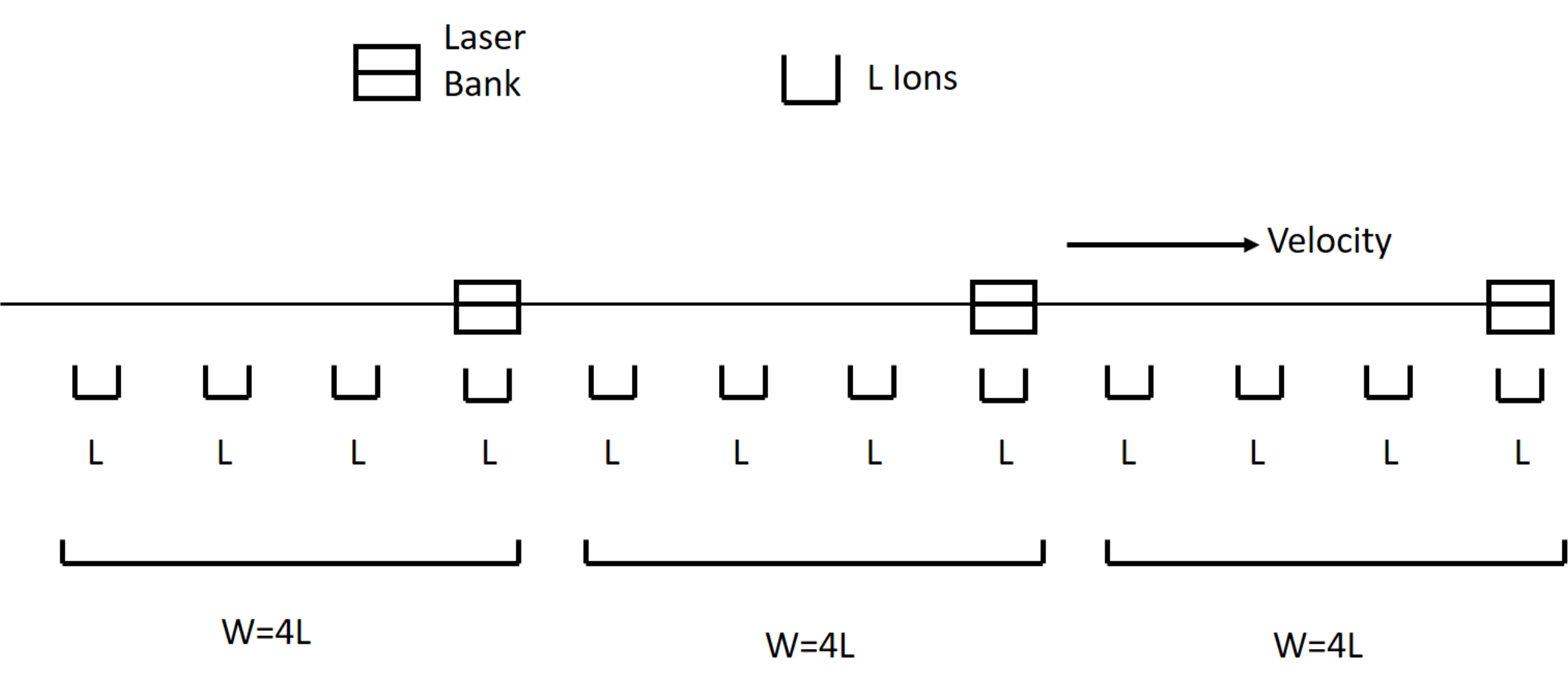}}
	\caption{Serial Mode of Operation, W=4L.}
	\label{fig-series-system}
\end{figure}

Figure 4 illustrates the details of this serial mode of operation (when sub-circuits do not overlap - see section 6).  In the left diagram the integer numbers indicate processing order of the sub-circuits of the circuit.  Sub-circuit 1 is processed first by the left most laser bank.  Once sub-circuit 1 rotates to the next laser bank, sub-circuit 2 is processed by the first laser bank and sub-circuit 1 is processed in parallel by the second laser bank in more depth (going left to right).  The shifting continues until the first four sub-circuits are being processed by the first four laser banks (top row in left diagram).

\begin{figure}[t]
	\centerline{\includegraphics[width=13cm, height=8cm]{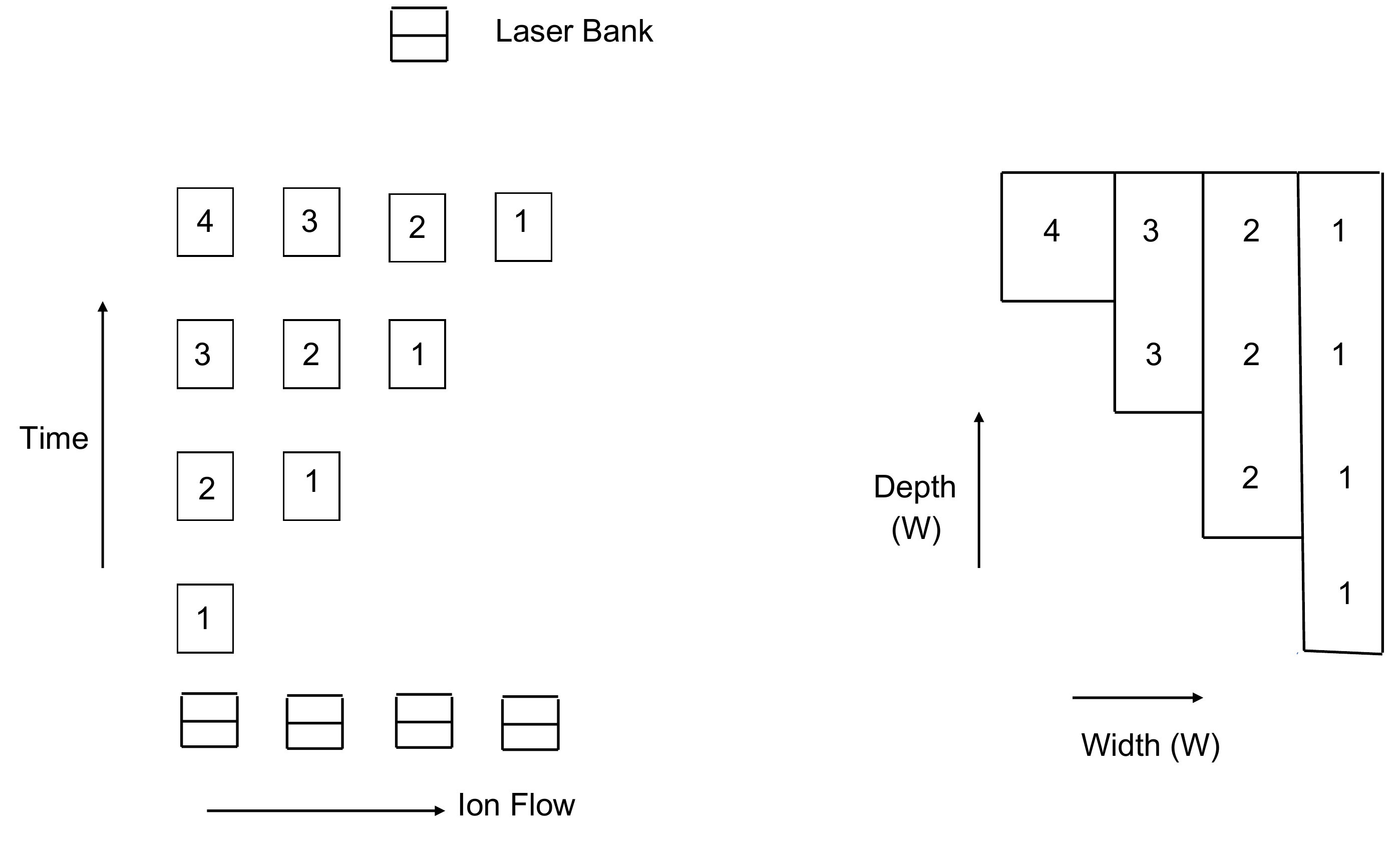}}
	\caption{Left Figure: Four Sub-Circuit Parts Being Fed Serially Thru a Number of Laser Banks.  Right Figure: 
		Implemented Circuit Topology.}
	\label{fig-SerialLayout2}
\end{figure}

Once all four sub-circuits are under laser banks, the $i$th sub-circuit will have been processed $i$ times in a depth wise sense (see right diagram in Figure 4).  Thus, initially at least, only triangular circuit topologies can be supported.  Once all four sub-circuits are under lasers further processing will be in more of a parallel sense (see parallel mode below) resulting in a trapezoidal topology circuit.  

Note that in comparing Figure 3 and Figure 4, Figure 3 has gaps of three bunches of ions between laser banks and Figure 4 does not.  Gaps could have been inserted into Figure 4 also - they were not to make the figure clearer.  Essentially Figure 4 could be thought of showing the programming of ions in every fourth bunch in Figure 3.

Some potential issues are: 

\vspace{0.1 in}

(1) Triangular topologies may have niche applications but are probably of less usefulness than general rectangular circuit topologies (see parallel mode discussion below).  

(2) The limit on the width of the circuit that could be processed is the number of ions in a bunch (L) times the number of bunches that 
can be  utilized for a circuit (B) or BL.

(3) Another serious issue is a gate in the interior of the circuit may have two quantum qubit inputs, say.  Each one may be from a preceeding sub-circuit processed by a different laser bank.  But in this serial mode of operation there may be delay (related to the rotation velocity) between when each input is available.  We refer to this as the predecessor problem.  We return to this issue in section 8.  

\vspace{0.1 in}

{\bf (B) Parallel Mode of Operation:}  Suppose there are L ions per laser bank in series along the SRQC.  Let the total width (in ions) of the circuit be W.  Each laser bank processes W/L ions concurrently.  They would process the first stage (depth wise), then the second stage and so on… (see Figure 5).   This could be done to a certain depth until the ions rotate away, perhaps to another set of laser banks (including some of the original ones) to continue the 
processing in an increased circuit depth sense.  The overall circuit topology is rectangular (consider serial mode without the initial triangular circuit topology).

This is illustrated in Figure 5.  In the diagram W=4L (there are four bunches of ions in the circuit width W).  Four laser banks (in the same or 
different windows) are shown programming a width 4L circuit to some depth.  Again the ions are uniformly spaced.  

If more depth wise processing is required, the ions can rotate to a successive group of laser banks (which may either partially overlap the original group of laser banks or not overlap the original group of laser banks).

\begin{figure}[t]
	\centerline{\includegraphics[width=11cm]{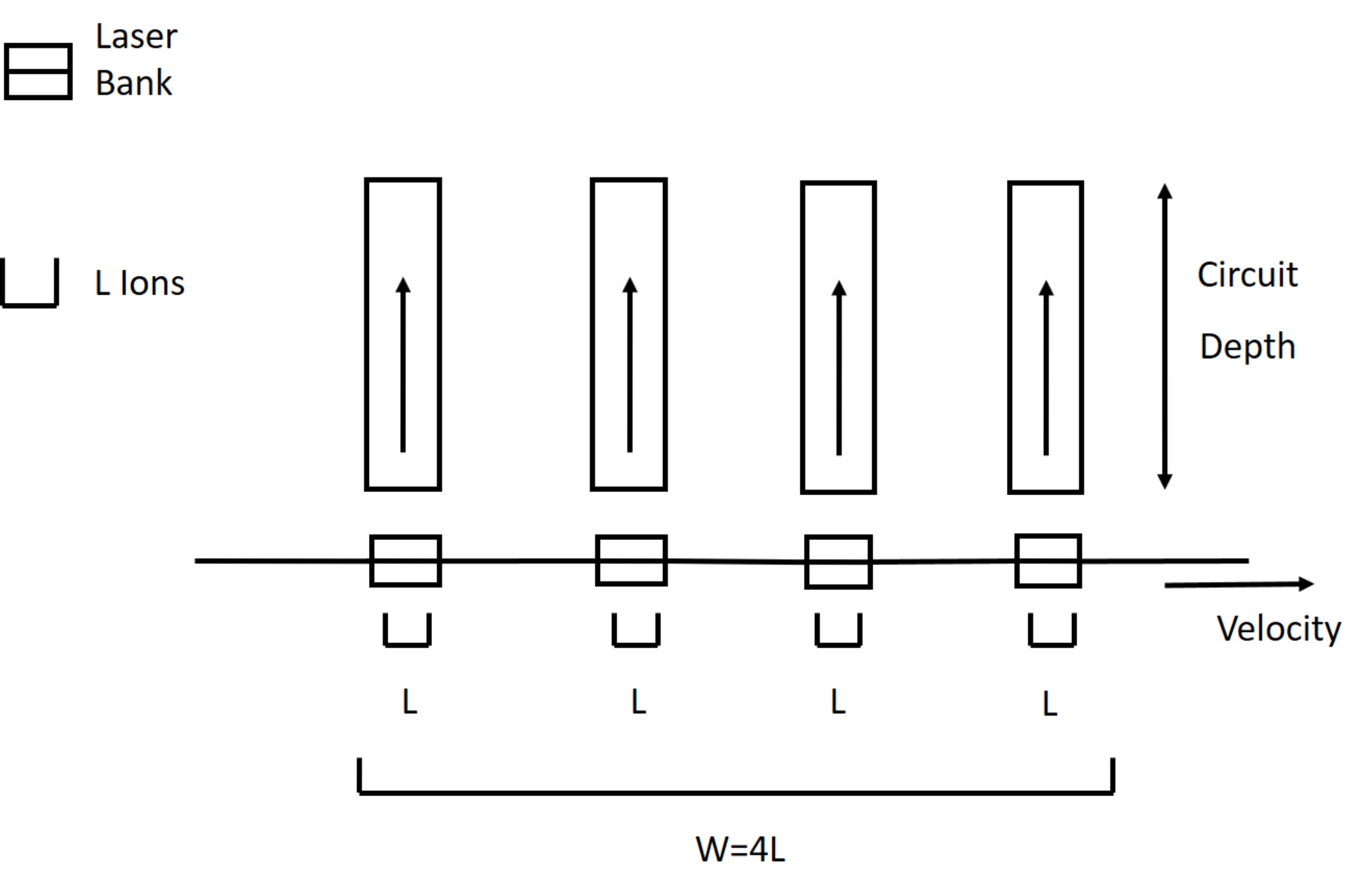}}
	\caption{Parallel Mode of Operation, W=4L.}
	\label{fig-parallel-system}
\end{figure}   

Some potential issues are:

\vspace{0.1 in}

(1)  There may be non-programmable sections of the SRQC between each adjacent          
pair of laser banks/windows.  Thus each set of adjacent programmed qubits is separated   
from the next by a group of what could be “dark” ions/qubits.  This may not necessarily 
be a problem but:

(2) A gate on the interior of the overall quantum circuit may have two quantum qubit 
inputs, say.  Each input may be from sub-circuits operated on by different laser 
banks.  Both inputs thus may not be available at the same time.   
This is a predecessor problem, again. How are the two inputs brought to the gate?  
This is addressed in section 8.  

\vspace{0.1 in}

{\bf (C) Hybrid Modes of Operation:}  If the predecessor problem can be addressed, it should be possible to have a mix of serial and parallel modes of operation.  In this section it was discussed how this can result in a trapezoidal circuit topology.  

\vspace{0.1 in}

\section{Timing Equations}

Referring to Figure 6, in terms of the timing of the processing let again $D$ be the overall depth and let $W$ be the overall width of the (native gate) circuit to be implemented.  Let $W^*$ 
be the maximal width and let $D^*$ be the maximal depth of the part of the circuit (sub-circuit) that a laser bank can implement in one pass.

\begin{figure}[t]
	\centerline{\includegraphics[width=14cm, height=10cm]{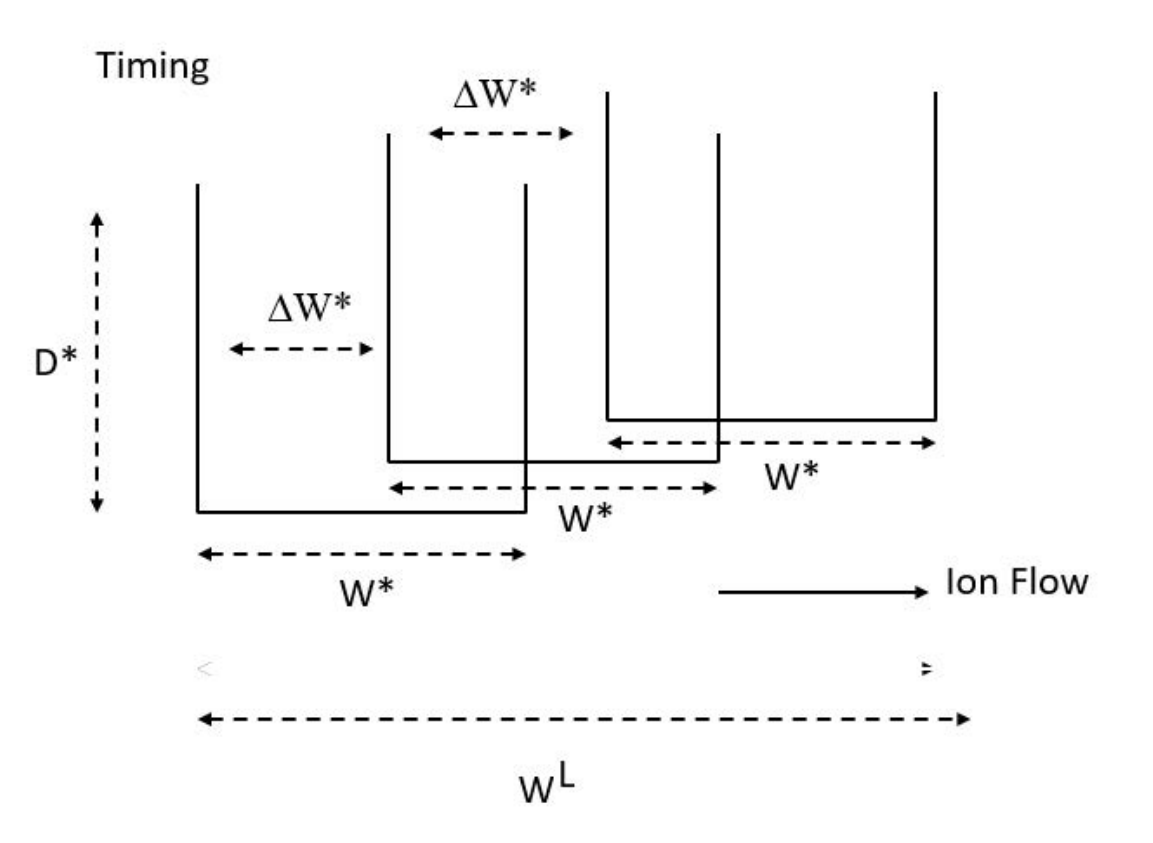}}
	\caption{Generic Timing Diagram.}
	\label{fig-Timing}
\end{figure} 

Three cases are now examined:

{\bf First Case:} Consider a serial mode case where the entire circuit width, $W$, is contained within $W^*$ (only one sub-circuit).  The circuit is brought under the first laser bank and programmed to as much depth $D^*$ as possible (Figure 6).  Then as the ions rotate around the storage ring they slide to the right in the diagram by amount $\Delta W^*$ and are programmed by possibly some of the same lasers and new lasers from the next bank to an additional depth $D^*$ each iteration.  

The total number of passes needed to map a $D$ depth circuit is then $\lceil \frac{D}{D^*} \rceil$. Here $\lceil x \rceil$ is the smallest integer greater than or equal to $x$.   Let $W^L$ be the \lq\lq width of the laser banks" needed to do this.

So the circuit width $W^*$ comes into view of the 
lasers first followed by $( \lceil \frac{D}{D^*} \rceil - 1 )$ shifts of the circuit of $\Delta W^*$ each.  A timing equation that finds $W^L$ in terms of the parameters is:

\begin{equation}
	W^L = W^* + \left( \lceil \frac{D}{D^*} \rceil - 1 \right) \Delta W^*
	\end{equation}

Again, this is the result where the entire circuit width, $W$, is contained within $W^*$ (only one sub-circuit).  

{\bf Second Case:} Now let $\Delta W^* = W^*$ (serial case).  That is, a sub-circuit (only one in this case again) of width $W^*$ is shifted completely each iteration by $W^*$ (no overlap).  Then by substituting $\Delta W^* = W^*$ into equation (1), the width of the laser banks  being used ($W_L$) is:  

\begin{equation}
	W^L = W^* \lceil \frac{D}{D^*} \rceil 
\end{equation}
	
{\bf Third Case:} Finally in parallel mode (section 5) in the first pass the initial width of the laser banks used ($W_L$) is $W$.  Then each time the circuit is slid by one laser bank the width of the laser banks used is increased by 
$W^*$.  Thus:

\begin{equation}
	W^L = W + ( \lceil \frac{D}{D^*} \rceil  - 1) W^*
\end{equation}

An alternative technique for a large depth circuit, is to transfer qubits whose time under the lasers is about up to locations that will come under the lasers/window next using swap gate(s) operations or teleportation.  The physical means of doing this is an open problem.  That way the entire circuit may be implemented at one set of lasers.   This would extend the depth of the circuit as additional qubits are brought in to serve to continue the calculations of one circuit.  This technique can be used to implement cascades of different quantum circuits which overall would constitute a very large depth quantum \lq\lq circuit".   See section 10 for some representative sizing and timing numbers.  

\section{Mismatched Inputs$/$Outputs}

Another case is

\vspace{0.15 in}

$\bullet$ Assumption 2: Some or all of the gates in a circuit have at least one output qubit location that is different from the input qubit locations of 
that gate.  
 
\vspace{0.15 in}
 
In this case this can be handled as above if the output qubits of the gate(s)$/$circuit fit within the width of the circuit input qubits.  
 
If some or all gate(s)$/$circuit outputs are not within the original circuit input qubit width, with the right storage ring and quantum computing timing parameters based on storage ring quantum computing physical layout, at least some of the gate(s)$/$circuit output qubits can be placed in upstream or downstream qubits for potential further use either at the same window or a different window.  The details of implementing this is an open problem.

\vspace{0.1in} 

\section{The Predecessor Problem}

The predecessor problem may occur in serial, parallel or hybrid implementations of large width circuits.  This is the problem that occurs 
when a gate has, as an example, two inputs and the two inputs are programmed from different laser banks.  

For the serial mode of operation the 
L ions are processed as a unit.  If a gate receives one input but doesn't immediately receive the second input, the gate can wait until the 
next bunch of L ions comes into view of the laser bank so that the second input can be supplied.  Ion states are long lasting 
in an ion trap or storage ring (on the order of minutes) so the first input qubit can wait though 
further circuit operation beyond the gate may have to wait also.

The predecessor problem can also appear in parallel implementations of large width circuits.  That is, a gate in a sub-circuit (involving L ions) 
may have, as an example, two inputs, only one of which is accessible to a laser bank because the second input is programmed by a different laser 
bank.  Again, in general because ion states are long lasting, a gate input(s) can wait for another input(s) to be supplied.   

Hybrid implementations of series and parallel configurations can also use delayed inputs as a means of solving the predecessor problem.

It is an open problem to develop software to pre-process circuits to be implemented on a SRQC by adjusting circuit and sub-circuit width, depth and layout in order to minimize/eliminate instances of the predecessor problem. See section 9A.

\section{Extensions}

Some other architectural options include:

{\bf (A) Software:}  It should be possible but is an open problem to develop software that pre-processes quantum circuits and specifies SRQC modes of operation and parameters (such as width and depth of each sub-circuit processed) to minimize the instances of the predecessor problem.  It could also adjust physical circuit versus logical circuit layout (that is adjust the physical layout while preserving the logical connectivity).  It should be noted that in a similar spirit software for pre-processing quantum circuits for efficient implementation has been developed for such purposes as mapping circuits to the IBM QX architecture 
\cite{Almeida, Kole, Rahman, Zulehner}.  

\vspace{0.1 in}

{\bf (B) Wrap-around:}	If the predecessor problem can be addressed, these modes of operations can be implemented in a “wrap-around” fashion.  This would allow circuit widths greater than the number of qubit ions that can be accommodated around the circumference.    

\vspace{0.1 in}

{\bf (C) Block by Block vs. Continuous Operation:}  Much of the discussion above involves laser 
banks processing \lq\lq blocks" (i.e. concatenations) of L ions at a time.  However it is possible to process ions continuously.  
That is, a laser may interact with an ion as soon as it comes in view of, say, a window rather than wait for all L ions in its block to come in view 
of the laser (or laser bank). 

\vspace{0.1 in}

{\bf (D) Logical and Physical Qubits:} There have been proposals to represent each \lq\lq logical" qubit by a number of \lq\lq physical" 
qubits/ions for purposes of reliable computation \cite{Kiel}.  The details of implementing this in a SRQC is an open problem but it fits into the methodology presented here with lasers accessing multiple physical qubits.

\vspace{0.1in}

\section{Representative SRQC Sizing and Timing}
	
Consider some representative parameters relating to the SRQC sizing and timing.  Assume a 1 meter in circumference ring.  Assume a window is 1 mm wide (it could be larger) and the ions have a 10 micrometer spacing (100,000 ions around the ring).  There will thus be 100 ions \lq\lq visible" to the lasers as they pass the window.

If the ions are traveling at a speed somewhere from 100 to 1000 m/s, a single ion will spend 10 to 1 microseconds going past the window (that is time = window size/speed).  

Suppose we can do a set or measurement or excite a phonon state in about a 1 to 10 microsecond time frame.  The time to write a qubit state is on the order of 1 to 10 microseconds from numerous publications and calculations based on C.J. Foot's book Atomic Physics \cite{Foot}.      The time to measure a qubit state is more complex and is at least of the same order but ongoing work may allow qubit state measurements down to tens of nanoseconds.

So it may be possible to do ten 1 microsecond \lq\lq actions" on a single 10 microsecond pass but probably only implement quantum gates of less than a depth of ten since a single gate may take several actions.  However, as discussed above, multiple windows or several passes around the ring (since ions are expected to be long lasting) would allow larger depth circuits.

A challenge is a tradeoff between ring latency and window access time.  With a 1 meter in circumference ring and a ion velocity of 100 m/sec (used in the previous paragraph) the time for ions to move completely around the ring and come back to an original window locations is .01 sec or 10 msec or 10,000 microseconds.  This seems like a large latency overhead for a 10 microsecond pass past a window.  If the velocity is increased to 1000m/sec the ring circumnavigation time is 1,000 microseconds, still appreciable.  Moreover the time to do a window pass now drops to 1 microsec, possibly not allowing appreciable laser operations during the pass (see paragraph above).

Of course additional windows around the ring would decrease latency to some extent.  If the write time can be improved by a factor of ten to about 0.1 micro seconds then an appreciable number of operations may fit into a window pass allowing a number of stages of gates to be programmed.

\section{Conclusion}

What has been presented are original architectural options for quantum circuit mapping to a storage ring quantum computer.  
Contributions of this work include:

\vspace{0.15 in}

$\bullet$ Series, parallel and hybrid arrangements and timing of circuit mapping.

$\bullet$ Examining other features that may play a role in quantum circuit mapping including 
circuit optimization software, wrap-around operation, block by block versus continuous 
operation and using multiple physical qubits to represent logical qubits.  

$\bullet$ Timing equations for serial and parallel cases.  

$\bullet$ A discussion of the predecessor problem and its resolution.

$\bullet$ Back of the envelope calculations of the sizing and timing of a SRQC.   

$\bullet$ Half a dozen largely engineering, not conceptual, open problems to implement a working SRQC are identified.  

\vspace{0.15in}

Storage ring quantum computers are a futuristic computing approach.  The methods 
presented here for circuit mapping onto a SRQC are general.  The SRQC architecture and circuit mapping proposed here embodies much parallelism in terms of ions, windows and lasers.  This all sets a foundation for future work and many open problems.

\section{Acknowledgements}

This work was performed under Contract No. DE-SC0012704 with the auspices of the U.S. Dept. of Energy.  Thank you to Dai Weining for a careful reading of the manuscript.

\end{document}